\documentclass[conference,a4paper]{APSIPA2021}
\let\labelindent\relax
\usepackage{amsmath}
\usepackage{graphicx}
\usepackage{multirow}
\usepackage{threeparttable}
\usepackage[utf8]{inputenc}  
\usepackage[backend=biber,style=ieee,]{biblatex}
\usepackage{color}
\usepackage{amsmath,amsfonts}
\usepackage{algorithmic}
\usepackage{multirow}
\usepackage{latexsym}
\usepackage{newtxtext,newtxmath}
\usepackage{algorithm}
\usepackage{enumitem}
\usepackage {diagbox}
\usepackage{here}
\usepackage{graphicx}
\usepackage{textcomp}
\usepackage{xcolor}
\usepackage{subcaption}
\usepackage{caption}
\usepackage{float}
\usepackage{makecell}
\usepackage{graphicx}
\usepackage{subcaption}
\usepackage{caption}

\usepackage{geometry}
\usepackage{fancyhdr}

\fancypagestyle{firststyle}{
  \fancyhf{}
  \fancyhead[C]{2024 Asia Pacific Signal and Information Processing Association Annual Summit and Conference (APSIPA ASC)}
}
\begin{document}
\title{Disposable-key-based image encryption for collaborative learning of Vision Transformer}

\author{
\authorblockN{
Rei Aso\authorrefmark{1},
Sakaya Shiota\authorrefmark{2} and
 Hitoshi Kiya\authorrefmark{3}
}

\authorblockA{
\authorrefmark{1}
Tokyo Metropolitan University, Tokyo \\
E-mail: aso-rei@ed.tmu.ac.jp \\
}

\authorblockA{
\authorrefmark{2}
Tokyo Metropolitan University, Tokyo \\
E-mail: sayaka@tmu.ac.jp}

\authorblockA{
\authorrefmark{3}
Tokyo Metropolitan University, Tokyo \\
E-mail: kiya@tmu.ac.jp}
}

\maketitle
\thispagestyle{firststyle}
\pagestyle{fancy}
\nocite{*}
\begin{abstract}
  We propose a novel method for securely training the vision transformer (ViT) with sensitive data shared from multiple clients similar to privacy-preserving federated learning. In the proposed method, training images are independently encrypted by each client where encryption keys can be prepared by each client, and ViT is trained by using these encrypted images for the first time. The method allows clients not only to dispose of the keys but to also reduce the communication costs between a central server and the clients. In image classification experiments, we verify the effectiveness of the proposed method on the CIFAR-10 dataset in terms of classification accuracy and the use of restricted random permutation matrices. 
\end{abstract}

\section{Introduction}
Deep neural networks (DNNs) have been deployed in various applications. Training a high-quality DNN model requires a huge amount of training data, but model training raises privacy concerns, especially when dealing with sensitive or personal information \cite{9934926}. To collect a huge amount of training data, in centralized machine learning, each client prepares information on local data and sends it to a central server for model training. Federated learning (FL) \cite{pmlr-v54-mcmahan17a, 9919201} is a learning approach that allows multiple clients to collaboratively train a shared model while keeping their data decentralized and private. In FL, clients participate in the training process using their local data, without sharing their raw data. However, the iterative learning process in FL generates some issues in terms of learning costs for clients and communication costs in a distributed environment \cite{zhang2021survey}. To overcome these issues, we propose a novel approach in which multiple clients efficiently collaborate to train a high-quality model.

In privacy-preserving collaborative learning, each client collects data and securely sends the data or an update obtained from the data to a central server for model training. In FL, in each round, each client independently computes an update to the current model based on its local data and communicates this update to a central server so that the client-side updates are aggregated to compute a new global model. Accordingly, conventional collaborative learning including FL requires both resources for computing the update in each client and communication efficiency \cite{9934926}. In this paper, we propose a novel method for privacy-preserving collaborative learning to reduce these costs. 
 The method is carried out by using encrypted data, called learnable encryption \cite{SIP-2021-0048} where data is independently encrypted by each client. The encrypted data is sent to a central server for model training one time only, and a global model is trained by using the encrypted data from clients, so the method allows us not only to reduce the communication costs but to also avoid the cost for computing an update at all clients. We make the following contributions in this paper.
 \begin{itemize}
\item[(a)] We propose a novel method using learnable encryption for privacy-preserving collaborative learning that can reduce both communication costs and computation sources of clients.
\item[(b)] We propose the use of restricted random permutation matrices to reduce the influence of data encrypted with independent keys.
\end{itemize}

In the proposed method, different random permutation matrices used as encryption keys are assigned to each piece of data, but this causes the performance of models to degrade. To improve this issue, we also consider the use of restricted random permutation matrices. In experiments, the effectiveness of the method is verified in an image classification task under the use of the vision transformer~(ViT) \cite{ViT}.

\section{Related Work}\label{Related work}
The proposed method will be discussed under the use of ViT and learnable image encryption, which are summarized here.
\subsection{Vision transformer}
\begin{figure}[bth]
    \centering
    \includegraphics[bb=0 0 806 500,scale=0.4]{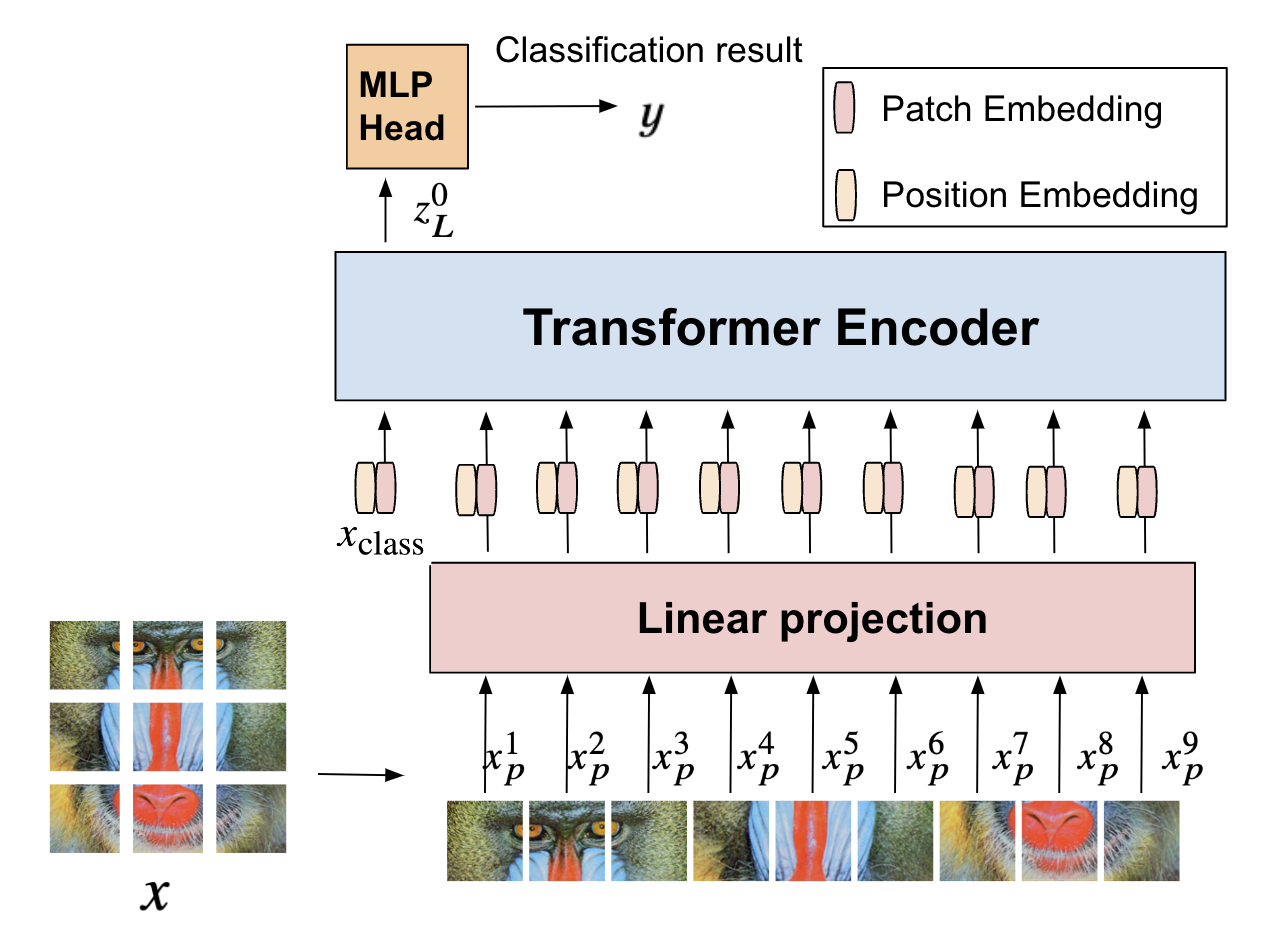}
    \caption{Overview of ViT}
\end{figure}
ViT \cite{ViT} is commonly used in image classification tasks and is known to provide a high classification performance. As shown in Fig. 1, in ViT, an input image $x \in \mathbb{R}^{h \times w \times c}$ is divided into $N$ patches with a size of $p \times p$, where $h$, $w$, and $c$ are the height, width, and number of channels of the image. Also, $N$ is given as $hw/p^2$. Afterward, each patch is flattened into $x_{p}^i = [x_{p}^i(1), x_{p}^i(2),\dots , x_{p}^i(L)]$. Finally, a sequence of embedded patches is given as
\begin{align} \label{original_input}
    z_{0} =& [x_{class}; x_{p}^1\mathbf{E}; x_{p}^2\mathbf{E}; \dots x_{p}^i\mathbf{E}; \dots x_{p}^N\mathbf{E}] + \mathbf{E_{pos}},
\end{align}
where 
\begin{align*}
\mathbf{E_{pos}} =& ((e_{pos}^0) (e_{pos}^1) \dots (e_{pos}^i) \dots (e_{pos}^N)),\\
  L =& p^{2}c, x_{class} \in \mathbb{R}^{D}, \ x_{p}^{i} \in \mathbb{R}^{L}, \ e_{pos}^i \in \mathbb{R}^{D},\\
  \mathbf{E} \in& \mathbb{R}^{L \times D}, \ \mathbf{E_{pos}} \in \mathbb{R}^{(N+1) \times D}.
\end{align*}
$x_{class}$ is a class token, $\mathbf{E}$ is an embedding (patch embedding) that linearly maps each patch to dimensions $D$, $\mathbf{E_{pos}}$ is an embedding (position embedding) that gives position information to patches in the image, $e_{pos}^0$ is the position information of a class token, $e_{pos}^i$ is the position information of each patch, and $z_0$ is a sequence of embedded patches. Afterward, $z_0$ is input into the transformer encoder. The encoder outputs only the class token, which is a vector of condensed information on the entire image, and it is used for classification. 

Previous studies have indicated that when models are trained with images encrypted by one client, the performance of the models is degraded compared with models trained with plain images \cite{Privacy-Preserving-Image-Classification, 9909972}. If multiple clients collaboratively train models by using images encrypted with independent keys, the performance of the models will be more degraded than that of one client. Accordingly, we propose a novel method that can improve the performance degradation even when multiple clients collaboratively train a model by using images encrypted with independent keys. 
\subsection{Learnable image encryption}
 Various image transformation methods that use a secret key, often referred to as perceptual image encryption, have been studied so far for many applications \cite{SIP-2021-0048}. In this paper, we focus on learnable images transformed with a secret key, which have been studied for deep learning.
 Learnable encryption enables us to directly apply encrypted data to a model as training and testing data. Encrypted images have no visual information on plain images in general, so privacy-preserving learning can be carried out by using visually protected images. In addition, the use of a secret key allows us to embed unique features controlled with the key into images. Adversarial defenses \cite{adv-def, 10530249, 9190904} and access control \cite{iiji, Maungmaung_Kiya_2021} are carried out with encrypted data using these unique features. \par
 
A block-wise learnable image encryption method (LE) with an adaptation layer was introduced \cite{8448772} as the first learnable image encryption method, and then another encryption method, a pixel-wise encryption (PE) method that does not use any adaptation layer, was proposed \cite{8931606}. However, both encryption methods are not robust enough against ciphertext-only attacks, as reported in \cite{chang2020attacksimageencryptionschemes, 9410466}. To enhance the security of encryption, LE was extended to an extended learnable image encryption method (ELE) by adding a block scrambling (permutation) step and a pixel encryption operation with multiple keys \cite{madono2020blockwisescrambledimagerecognition}. However, ELE still has inferior accuracy compared with using plain images, even when an additional adaptation network is used to reduce the influence of the encryption. Recently, block-wise encryption was also pointed out to have a high similarity to isotropic networks such as ViT and ConvMixer \cite{9909972, jimaging8090233}, and the similarity enables us to reduce performance degradation. However, no conventional learnable encryption methods have been designed for privacy-preserving model training with multiple clients. 

\section{Proposed Method}\label{Proposed Method}
\subsection{Overview}
Fig. 2 shows an overview of the proposed method, where datasets that M clients have are shared to train a global model. In the framework, a pretrained model is fine-tuned by using the shared datasets. Below is the procedure for model training and testing.
\begin{quote}
 \begin{itemize}
  \item[1] Each client prepares keys for image encryption and carries out image encryption with the keys.
  \item[2] Each client sends the encrypted images to a central server.
  \item[3] A global model is fine-tuned by using the encrypted images on the server. 
  \item[4] Each client receives the trained model from the server.
  \item[5] Each client encrypts a test image by using a key and inputs it to the model to obtain an estimation result. 
 \end{itemize}
\end{quote}
In this framework, the server cannot get both visual information of images and the keys. In addition, each client can use a key that is different from other clients' keys. Clients may also change keys for each image.
\begin{figure}[bth]
    \centering
    \includegraphics[bb=10 0 846 520,scale=0.45]{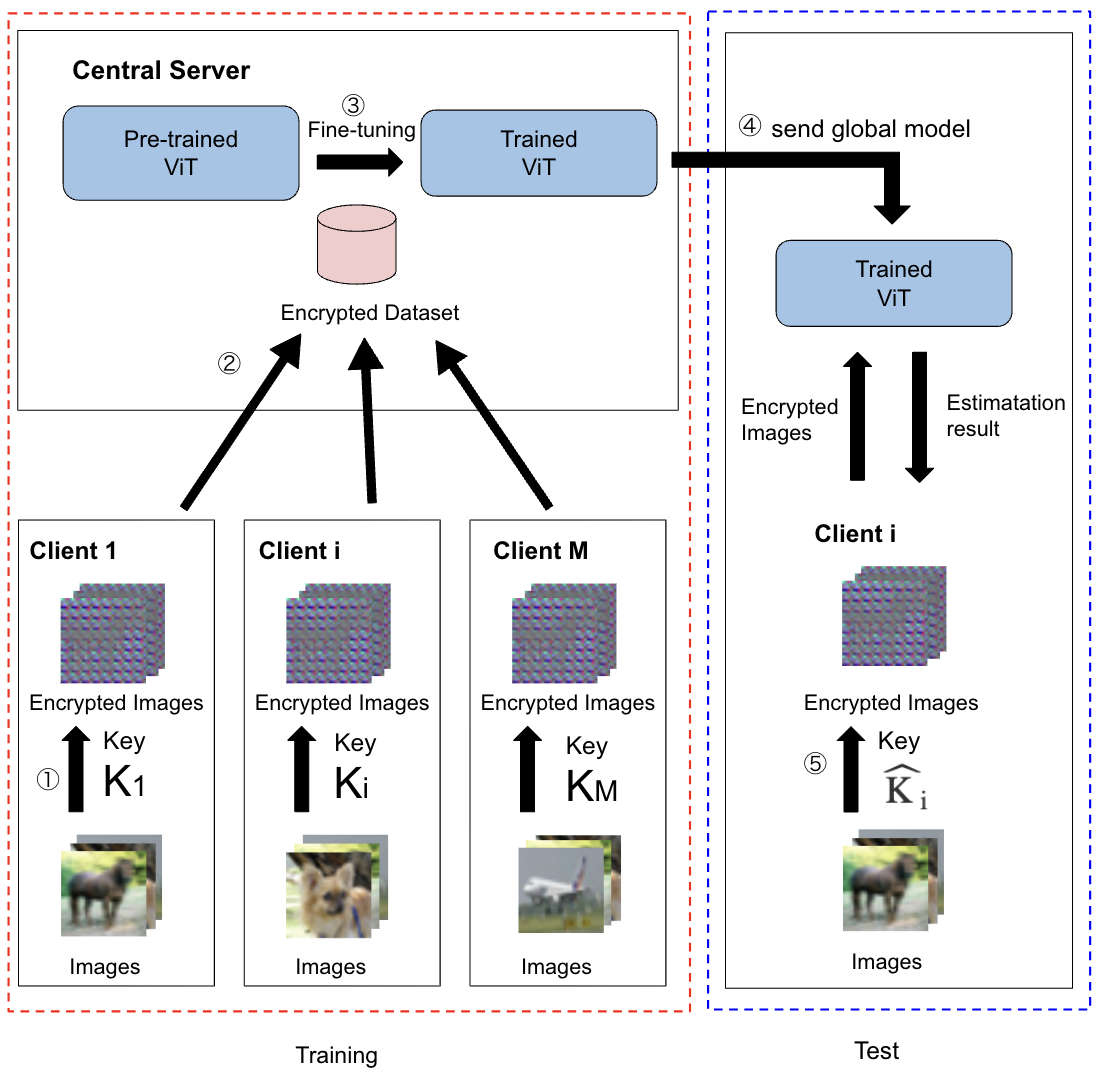}
    \caption{Framework of proposed method}
\end{figure}
\subsection{Image encryption with random permutation matrices}
ViT has a high similarity to block-wise encryption \cite{Privacy-Preserving-Image-Classification}, so the proposed method is carried out using block-wise encryption, which consists of block scrambling (block permutation) and pixel permutation. Below is the procedure for block scrambling, in which $\bf{E}_{bs}$ is a random permutation matrix for block permutation, given by
\begin{gather}
\bf{E}_{bs} = \begin{pmatrix} 
  E_{bs}(1,1) & \dots & E_{bs}(1,j) & \dots & E_{bs}(1,N) \\
  \vdots &       & \vdots &       & \vdots \\
  E_{bs}(i,1) & \dots & E_{bs}(i,j) & \dots & E_{bs}(i,N) \\
  \vdots &       & \vdots &       & \vdots \\
  E_{bs}(N,1) & \dots & E_{bs}(N,j) & \dots & E_{bs}(N,N)
\end{pmatrix} .
\end{gather}
\begin{itemize}
\item[(a)] Divide an image $x \in \mathbb{R}^{h \times w \times c} $ into $N$ non-overlapped blocks with a size of $p \times p$ such that $\textbf{B} = \{ B_{1},B_{2},...,B_{N}\}^\top $,~$B_i \in \mathbb{R}^{p^2c}$, where $p$ is also equal to the patch size of ViT.
\item[(b)]	Generate a random integer sequence with a length of $N$ by using a key as
    \begin{gather}
      l = [l(1), \cdots l(i), \cdots l(N)] \\
      \hspace{-200pt} \text{where} \nonumber \\
      l(i) \in \{1,2, \cdots, N\}, \nonumber \\
       l(i) \neq l(j)~~\text{if}~~ i \neq j,~~i,j \in \{1,2, \cdots, N\}. \nonumber
    \end{gather}
\item[(c)] Define $\bf{E}_{bs}$ in Eq.(2) as
    \begin{gather}
      E_{bs}(i,j) = 
        \begin{cases}
        0 & (l(j)\neq i) \\
        1 & (l(j) = i)
        \end{cases}.
    \end{gather}
\item[(d)] Define a vector b as 
    \begin{gather}
     b= \{ 1,2, \cdots, N \},
    \end{gather}
and transform it with  $\bf{E}_{bs}$ as
    \begin{gather}
      \hspace{-150pt} \hat{b} = b\bf{E}_{bs}  \\
      \hspace{15pt} = \{ \hat{b}(1),..,\hat{b}(i) .., \hat{b}(N) \}, ~~ \hat{b}(i) \in \{ 1, 2, .., N\}. \nonumber
    \end{gather}
\item[(e)] Give permutated blocks $\hat{\textbf{B}} = \{ \hat{B}_{1},\hat{B}_{2},...,\hat{B}_{N}\}^\top $ as
    \begin{gather}
     \hat{B_i} = B_{\hat{b}(i)}.
    \end{gather}
\end{itemize}
\par
In the proposed method, each client carries out the above procedure with their keys.
 Next, the procedure for pixel permutation is explained. In the method, R, G, and B values in each block are randomly permutated. A random permutation matrix for pixel permutation $\bf{E}_{ps}$ , which is applied to each block, is expressed as
\begin{gather}
\bf{E}_{ps} = \begin{pmatrix} 
  E_{ps}(1,1) & \dots & E_{ps}(1,j) & \dots & E_{ps}(1,L) \\
  \vdots &       & \vdots &       & \vdots \\
  E_{ps}(i,1) & \dots & E_{ps}(i,j) & \dots & E_{ps}(i,L) \\
  \vdots &       & \vdots &       & \vdots \\
  E_{ps}(L,1) & \dots & E_{ps}(L,j) & \dots & E_{ps}(L,L)
\end{pmatrix} ,
\end{gather}
    \begin{gather}
      \hspace{-200pt} \text{where}~~~ L = p^2C. \nonumber 
    \end{gather}
Below is the procedure for the permutation.
\begin{itemize}
\item[(a)] Generate a random integer sequence with a length of L by using a key as
    \begin{gather}
       u = [u(1), \cdots u(i), \cdots u(L)]. \\
      \hspace{-200pt} \text{where} \nonumber \\
       u(i) \in \{1,2, \cdots, L\}, \nonumber \\
       u(i) \neq u(j)~~\text{if}~~ i \neq j,~~i,j \in \{1,2, \cdots, L\}. \nonumber
    \end{gather}
\item[(b)] Define $\bf{E}_{ps}$ in Eq.(8) as
    \begin{gather}
      E_{ps}(i,j) = 
        \begin{cases}
        0 & (u(i)\neq j) \\
        1 & (u(i) = j)
        \end{cases}.
    \end{gather}
\item[(c)] Vectorize the elements of each block $B_i$ as $x_p^i \in \mathbb{R}^{L}$, and transform it with $\bf{E}_{ps}$ with
    \begin{gather}
      {x'}_p^i = x_p^i \textbf{E}_{\textbf{ps}}.
    \end{gather}
\item[(d)] Concatenate the transformed vectors into an encrypted image block $\hat{B}_i$.
\end{itemize}
\subsection{Use of restricted random permutation matrices}
A permutation matrix is a square binary matrix that has exactly one entry of 1 in each row and each column with all other entries 0. Every permutation matrix is orthogonal, with its inverse equal to its transpose. As describe above, two permutation matrices, $\bf{E}_{bs}$ and $\bf{E}_{ps}$, are used for image encryption in the proposed method. \par
Three types of permutation matrices are considered. The performance of trained models can be managed by the type of permutation matrix. For $N=5$, an example with $\bf{E}_{bs}$ is given below, where $^*$ indicates a fixed element.
\begin{itemize}
\item[(A)] Identity matrix $(N_{bs}=N)$:
\begin{gather}
\bf{E}_{bs} = \begin{pmatrix}
1^* & 0 & 0 & 0 & 0 \\
0 & 1^* & 0 & 0 & 0 \\
0 & 0 & 1^* & 0 & 0 \\
0 & 0 & 0 & 1^* & 0 \\
0 & 0 & 0 & 0 & 1^*
\end{pmatrix}.
\end{gather}
If $E_{bs}$ is the identity matrix of $N \times N$, in which all its diagonal elements equal 1, and 0 everywhere else, no permutation is carried out. In this type, the number of fixed diagonal elements $N_{bs}$ is given as $N_{bs}=N$.
\item[(B)] Restricted random permutation matrix $(0<N_{bs}<N)$: \\
$N_{bs} < N$ diagonal elements are fixed at a value of 1, where the positions of the fixed elements are randomly selected. When using $0<N_{bs}<N$, $\bf{E}_{bs}$ is called a restricted random permutation matrix. For $N_{bs}=1$,
\begin{gather}
\bf{E}_{bs} = \begin{pmatrix}
1^* & 0 & 0 & 0 & 0 \\
0 & 0 & 0 & 0 & 1 \\
0 & 0 & 1^* & 0 & 0 \\
0 & 1 & 0 & 0 & 0 \\
0 & 0 & 0 & 1 & 0
\end{pmatrix}.
\end{gather}
\item[(C)] Unrestricted random permutation matrix $(N_{bs}=0)$: \\
The positions of N elements are randomly selected as
\begin{gather}
\bf{E}_{bs} = \begin{pmatrix}
0 & 0 & 1 & 0 & 0 \\
1 & 0 & 0 & 0 & 0 \\
0 & 0 & 0 & 0 & 1 \\
0 & 1 & 0 & 0 & 0 \\
0 & 0 & 0 & 1 & 0
\end{pmatrix}.
\end{gather}
\end{itemize}
~~The above equations correspond to block scrambling (permutation) operations in each image. For $N_{bs}=N$, Eq. (13) is reduced to Eq. (12), and Eq. (13) is reduced to type (C) for $N_{bs}=0$. Similarly, $\textbf{E}_{\textbf{ps}}$ is also classified into three types, so $N_{ps} < L$ diagonal elements are fixed as in Eq. (13). The transformation with $\textbf{E}_{\textbf{ps}}$ is called pixel shuffling or pixel permutation. Fig. 5 shows an example of encrypted images, where $N=196$ and $L=768$ were used.

\subsection{Properties of proposed method}
The properties of the proposed method are summarized below.
\begin{itemize}
  \item Each client can share their images encrypted with their keys to train a global model.
  \item The performance degradation of models caused by encrypted images can be reduced by using the embedding structure of ViT and restricted permutation matrixes. 
  \item Transmission between clients and the central server is carried out to share encrypted images only one time during model training, so the communication cost of the proposed method is lower than conventional methods such as federated learning. 
\end{itemize}
 In addition to the above properties, each client can use different keys for each image. Accordingly, the keys used for image encryption are not required to be managed carefully. 
 \begin{figure}[ht]
    \vspace{-30pt}
    \centering
    \begin{subfigure}[b]{0.2\textwidth}
        \centering
        \includegraphics[bb=-30 -30 389 389,scale=0.4]{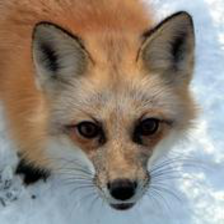}
        \caption{original}
        \label{fig:sub1}
    \end{subfigure}
    \hfill
    \begin{subfigure}[b]{0.2\textwidth}
        \centering
        \includegraphics[bb=-30 -30 389 389,scale=0.4]{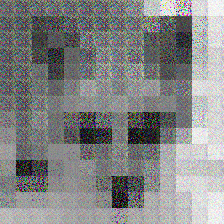}
        \caption{$N_{ps}=0$,~$N_{bs}=196$}
        \label{fig:sub2}
    \end{subfigure}

    \vspace{-3em}

    \begin{subfigure}[b]{0.2\textwidth}
        \centering
        \includegraphics[bb=-30 -30 389 389,scale=0.4]{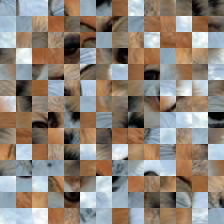}
        \caption{$N_{ps}=768$,~$N_{bs}=0$}
        \label{fig:sub3}
    \end{subfigure}
    \hfill
    \begin{subfigure}[b]{0.2\textwidth}
        \centering
        \includegraphics[bb=-30 -30 389 389,scale=0.4]{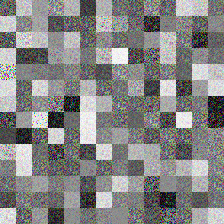}
        \caption{$N_{ps}=0$,~$N_{bs}=0$}
        \label{fig:sub4}
    \end{subfigure}
    \caption{Example of encrypted images}
    \label{fig:main}
\end{figure}
\begin{table}[h]
  \centering
  \caption{Classification accuracy(\%)}
\begin{tabular}{c|ccc}
   & w/o encryption & common key & different key \\
   &  &  &($N_{bs}=0$,~$N_{ps}=0$) \\
  \hline
   acc(\%) & 97.68 & 88.1 & 82.71 \\
\end{tabular}
\vspace{-30pt}
\end{table}
\begin{figure}[bth]
    \centering
    \includegraphics[bb=10 0 846 520,scale=0.35]{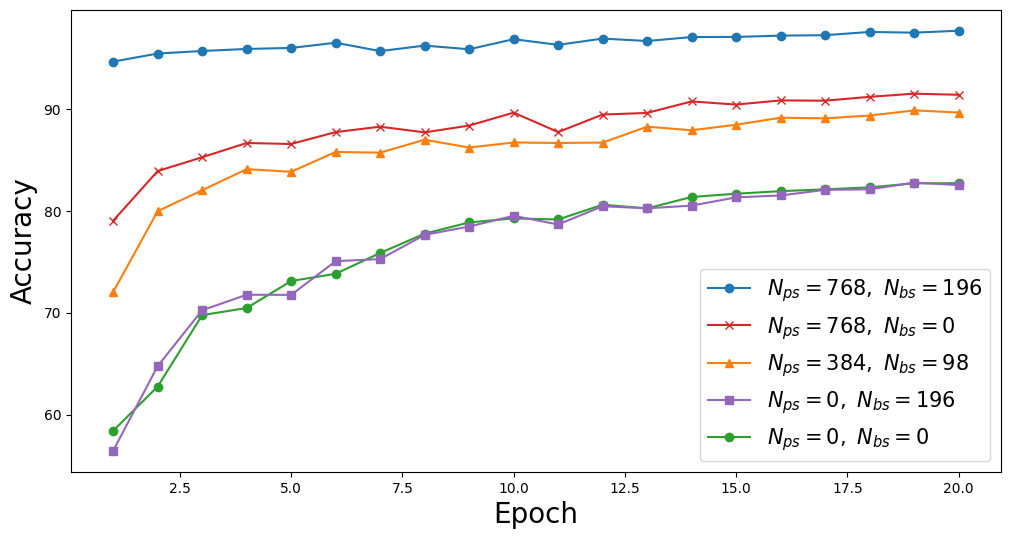}
    \caption{Learning curve of training process}
\end{figure}

\begin{table}[h]
  \centering
  \caption{Accuracy of using restricted matrix}
  \begin{tabular}{c|c}
    Image~($N_{bs}$,~$N_{ps}$) & Accuracy~(\%) \\
    \hline
    Baseline~(plain) & 97.02   \\
    Proposed~(0, 768) & 91.39   \\
    Proposed~(147, 576)  & 90.69  \\
    Proposed~(49, 576) & 90.66  \\
    Proposed~(98, 384) & 89.65 \\
    Proposed~(147, 192) & 88.33 \\
    Proposed~(49, 192) & 88.19 \\
    Proposed~(196, 0) & 82.53 \\
    Proposed~(0, 0) & 82.71 \\
  \end{tabular}
\end{table}

\begin{figure}[h]
    \vspace{-100pt}
    \centering
    \includegraphics[bb=10 0 846 520,scale=0.6]{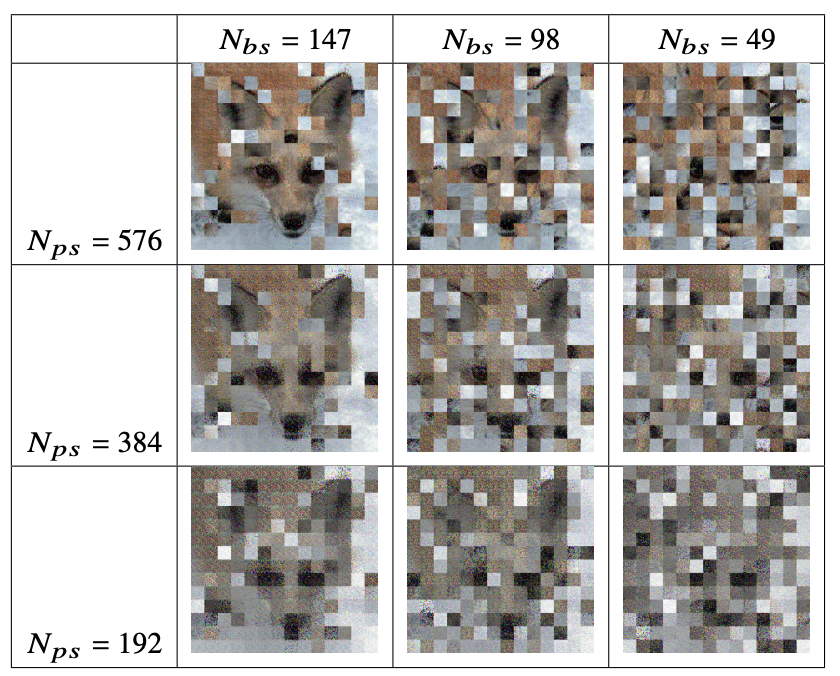}
    \caption{Example of encrypted images using restricted random permutation matrix}
\end{figure}
\vspace{30pt}
\section{Experiment Result}\label{Experiment Result}
The effectiveness of the proposed method was verified in experiments. 

\subsection{Setup}
In experiments, a model was prepared that was pre-trained trained with Image-Net, ``vit\_base\_patch16\_224," from the timm library, and the CIFAR10 dataset, which consists of $50,000$ training and $10,000$ test color images with a size of $32 \times 32$, was used to fine-tune the pre-trained model. Each client was given $10,000$ randomly selected images as training data without duplicates, where images were resized from $32 \times 32 \times 3$ to $224 \times 224 \times 3$ to fit the size of images to that of ViT. Fine-tuning was carried out under the use of a batch size of $64$, a learning rate of $0.0001$, a momentum of $0.9$, and a weight decay of $0.0005$ using the stochastic gradient descent (SGD) algorithm for $20$ epochs. A cross-entropy loss function was used as the loss function. We evaluated the classification accuracy by inputting the $10,000$ test images to the final global model. In the setting of ViT, the patch size p in patch embedding was set to $16$, the number of split patches in an input image was $N = 196$, and the dimensionality of output feature vectors was $D = 384$ and $L = 768$. In addition, independent keys were assigned to all images every epoch.  

\subsection{Classification accuracy}
In Table 1, the classification accuracy of the proposed method is compared with that of the baseline method without encryption to confirm the influence of encryption conditions, where common key indicates that all clients used the same keys, and different key means that independent keys generated with $N_{bs}=N_{ps}=0$, which were to use unrestricted random permutation matrices, were applied to all images. From the table, even when different keys were applied to images, the proposed method still allowed us to maintain a high classification accuracy. However, the accuracy of the models decreased due to the influence of image encryption. Table 2 shows the result of using restricted random permutation matrices. Fig. 3 shows the learning curves during training. From these results, the use of restricted random permutation matrices was demonstrated to be effective in improving the accuracy of the models. Some restricted conditions outperformed the method using a common key.  

\subsection{Visibility of encrypted images}
Fig. 5 shows an example of encrypted images under restricted conditions. From the figure, a less restricted condition gave encrypted images with higher visual information than the original ones. In contrast, with a less restricted condition, a model could be trained with a higher quality in general. Accordingly, restricted permutation matrices should be selected carefully on the basis of classification accuracy and visual protection. 
\subsection{Security analysis} 
In collaborative learning, the privacy of all local data has to be protected. In this framework, both the server that trains the global model and other clients are assumed to be untrusted, so they may try to restore original data from encrypted data. The objective of an attacker is to restore visual information from encrypted images. We assume that the attacker has access to encrypted images and the encryption algorithm but does not possess the secret keys. Accordingly, the attacker can only carryout ciphertext-only attacks (COAs) using encrypted images. Block-wise image encryption methods have been studied in terms of robustness against various COAs including state-of-the-art ones \cite{9410466, info14060311}. The proposed method allows clients to assign independent keys to each image. Therefore, when applying independent keys to images, encrypted images are more robust against state-of-the-art attacks than images encrypted with a common key.

\section{Conclusion}
In this paper, we proposed a novel method for collaborative learning with ViT. The method, which uses encrypted images, allows clients not only to assign different keys for encryption but to also use restricted random matrices to improve the accuracy of models. Compared with conventional methods such as FL, the communication costs and computation resources of clients can be reduced by using our method. In experiments, the effectiveness of the method was demonstrated in terms of accuracy and visual protection.
\printbibliography
\end{document}